\documentclass[conference]{IEEEtran}
\IEEEoverridecommandlockouts
\usepackage{cite}
\usepackage{amsmath,amssymb,amsfonts}
\usepackage{algorithmic}
\usepackage{graphicx}
\usepackage{textcomp}
\usepackage{xcolor}
\usepackage[table]{xcolor}
\usepackage{multirow}
\usepackage{hyperref}
\usepackage{xcolor}
\usepackage{subcaption}
\usepackage{booktabs}
\usepackage{tabularx}

\hypersetup{
  colorlinks=true,
  linkcolor=red,
  citecolor=red,
  urlcolor=red
}

\newtoggle{fullpaper}
\newtoggle{asia}
\newtoggle{shortpaper}

\newcommand{\full}[1]{\iftoggle{fullpaper}{#1}{}}
\newcommand{\shortpaper}[1]{\iftoggle{shortpaper}{#1}{}}

\newcommand{\asia}[1]{\iftoggle{asia}{#1}{}}

\toggletrue{fullpaper}
\togglefalse{shortpaper}
\toggletrue{asia}

\def\BibTeX{{\rm B\kern-.05em{\sc i\kern-.025em b}\kern-.08em
    T\kern-.1667em\lower.7ex\hbox{E}\kern-.125emX}}
\begin{document}

\title{How Global Is Anycast?: \\Evaluating Performance Gaps
}

\author{\IEEEauthorblockN{Mohammad Shaikh Sadid Khan}
\IEEEauthorblockA{\textit{Department of Computer Science} \\
\textit{California State University, Fullerton}\\
sadidkhan@csu.fullerton.edu}
\and
\IEEEauthorblockN{Karishma Setya}
\IEEEauthorblockA{\textit{Department of Computer Science} \\
\textit{California State University, Fullerton}\\
ksetya1@csu.fullerton.edu}
\and
\IEEEauthorblockN{ASM Rizvi}
\IEEEauthorblockA{\textit{Department of Computer Science}} 
\textit{California State University, Fullerton}\\
asmrizvi@fullerton.edu}

\maketitle

\begin{abstract}
Anycast networks aim to reduce latency by directing users to nearby service sites. However, users can still experience high latency even when accessing services via anycast. Some anycast deployments are localized and intended to serve a specific geographic region, while others are globally distributed to support users worldwide, with the expectation of consistently low latency across locations. In practice, latency can remain poor for certain users due to factors such as routing inefficiencies and peering relationships. These inefficiencies are often more pronounced in regions such as Africa\asia{, Asia} or South America, where Internet connectivity is generally less mature than in North America or Europe. Despite this, there has been no in-depth analysis of anycast performance in these continents. In this paper, we examine the latency experienced by users in these regions when accessing anycast networks. We analyze more than 10,000 anycast prefixes, quantify the severity of latency problems, and show that clients in Africa\asia{, Asia,} and South America experience substantially higher latency than those in North America. We also investigate the underlying causes of this performance gap.

\end{abstract}


\section{Introduction}
  \label{sec:intro}



In an anycast network, multiple sites provide the same service 
  by announcing the same service prefix from all sites~\cite{li2018internet}.
 Since multiple sites announce the same service prefix,  a user 
  usually 
  has multiple routing paths
  to different anycast sites.
  
Based on the Border Gateway Protocol (BGP)
  path selection algorithm,
  a single path is selected from multiple paths
  to a single anycast site~\cite{caesar2005bgp}.
The BGP path selection process is generally expected to route 
  users to the closest site, since—when other attributes like weights or local preferences 
  are equal—BGP prefers the shortest autonomous system path (AS-path).

Domain Name System (DNS) and Content Delivery Networks (CDNs) widely use anycast to serve clients from nearby locations by announcing the same prefix from multiple geographic sites. However, because BGP is not performance-aware and instead selects routes based on fixed attributes, anycast routing does not always deliver optimal latency~\cite{koch2021anycast}.

Network operators deploy both local and global anycast networks~\cite{zhou2023regional, koch2021anycast}. 
Local anycast is intended to serve nearby users, 
  while global anycast is designed to serve users worldwide.
However, in practice, 
  some anycast networks do not achieve this goal for various reasons.
First, Internet connectivity is generally more mature 
  in regions such as North America and Europe than 
  in many other continents, making it difficult 
  for operators to deliver comparable latency everywhere. 
Second, network administrators often prioritize deployments 
  in locations with larger user bases and greater revenue potential.
Third, undesirable routing behavior or peering relationships 
  can make it difficult for network operators to 
  provide comparable latency across all continents. 

Anycast polarization occurs when users in 
  the same geolocation experience very 
  different performance—some receive low latency 
  while others see high latency~\cite{rizvi2024anycast}. 
Prior work shows that this phenomenon is not uncommon 
  in regions such as North America and Europe. 
We hypothesize that polarization is even more severe 
  in other continents where Internet infrastructure 
  is less mature.
Also, 
  network operators deploy local anycast sites 
  intended to serve specific autonomous systems (ASes)\cite{li2018internet, koch2021anycast}. 
This can further increase the likelihood of polarization.

In this paper, we make three contributions. 
First, we show how ASes announce prefixes from different anycast sites (\autoref{sec:how-ases}). 
We demonstrate that a single AS can operate multiple anycast prefixes with markedly different performance. These cross-prefix differences highlight which prefixes are most important for clients in particular geographic locations (\autoref{sec:big-ases-latency}).
Second, to our knowledge, we are the first to quantify the degree of anycast polarization in regions such as Africa\asia{, Asia} and South America and show that polarization is especially pronounced in these continents (\autoref{sec:polarization}). In Africa, even under the best-case scenario, more than 10\% of probes experience latencies above 50\,ms.
Third, we examine anycast network configurations to identify the causes of polarization (\autoref{sec:reasons}). We find that large transit providers sometimes lack connectivity to nearby anycast sites, causing clients to be routed to distant sites and incur higher latency. We also characterize how large ASes and local peering relationships influence anycast performance along with a comparative analysis among large ASes.


\section{Related Work}
  \label{sec:related}
  
Prior work has evaluated anycast networks extensively. These studies examine when anycast outperforms unicast and when it leads to suboptimal site selection~\cite{calder2015analyzing}, identify routing-policy and route-selection effects that contribute to anycast inefficiencies~\cite{li2018internet}, and quantify how the number of anycast sites influences performance~\cite{de2017anycast}. Other work highlights the existence of regional anycast deployments that primarily serve clients within the same region~\cite{zhou2023regional}. In addition, prior studies demonstrate that performance can be improved through traffic-engineering techniques such as AS-path prepending to steer clients toward better-performing sites~\cite{zhou2026anypro}, as well as through operational configuration changes~\cite{mcquistin2019taming}. In contrast, our goal is to characterize anycast performance and configuration practices in regions such as Africa\asia{, Asia,} and South America, where Internet connectivity and interconnection ecosystems are comparatively less mature.

The impact of large ASes and peering relationships on latency has been widely studied. Prior work shows that anycast deployments with many upstream providers tend to exhibit more single-hop paths than deployments with only a few providers~\cite{mcquistin2019taming}, and that Internet Exchange Points (IXPs) can play a critical role in improving reachability and interconnection outcomes~\cite{prehn2022peering}. While earlier work characterized inter-domain connectivity in Africa in 2015~\cite{fanou2015diversity}, the region’s interconnection landscape has evolved substantially since then, motivating a renewed analysis focused specifically on anycast deployments. The work most closely related to ours identifies “polarization” effects arising from incomplete connectivity of large providers and local anycast sites~\cite{rizvi2024anycast}; however, it limits its analysis largely to North America and Europe, noting that polarization is even more severe in other continents. In contrast, we address this gap by analyzing the under-studied regions and by quantifying both the severity and underlying causes of polarization.

\section{Methodology}
  \label{sec:methodology}

We at first get the list of anycast addresses,
  and then make ping measurements from RIPE probes
  to compare latencies across continents.
We also make traceroutes to selective target prefixes
  to understand the underlying causes of polarization.
We use DNS logs to verify that real clients exist 
  behind the anycast prefixes.

\textbf{Getting anycast targets:}
We obtain the anycast addresses from a 
  prior study~\cite{manycast} that used active measurements to identify anycast IPs. 
To ensure high confidence in these addresses, 
  we follow their recommendations and 
  retain only those prefixes satisfying $(AB > 3) || (GCD > 1)$ value, 
  where AB means anycast based approach, and 
  Great-Circle Distance (GCD) is the latency based approach.

\textbf{RIPE measurements to the anycast targets:}
After compiling the anycast address list, 
  we use RIPE Atlas~\cite{ripeatlas} probes to measure latency to these anycast targets. 
We select at least \iftoggle{asia}{150}{250} randomly selected probes per continent and 
  run latency measurements to all anycast addresses, and 
  then compare results across continents.

To avoid bias in probe selection, 
  we choose probes from a diverse set of 
  autonomous systems and countries. 
Since probes within the same AS and country often exhibit 
  similar routing behavior, 
  we include at most two when multiple such probes are available.
Otherwise, we prioritize country diversity 
  within each continent to capture performance 
  across different geolocations.

We perform both ping and traceroute measurements 
  to the anycast targets. 
We first use ping to assess latency, and 
  when deeper insight into the network path is needed, 
  we run traceroutes to better understand 
  the underlying topology.

\textbf{Domain Name System (DNS) logs:}
To determine whether the anycast hostnames are queried 
  by clients in specific geographic regions, 
  we analyze DNS logs to identify where these hostnames are requested from. 
We use B-root server logs~\cite{usc_lander_ditl_broot_2026}, 
  which contain DNS query names and the corresponding 
  source IP addresses. 
We then geolocate the source IPs using IP-to-geolocation data
  from Geolite~\cite{maxmindgeolite} and ipinfo~\cite{ipinfo}.

\section{How Do Autonomous Systems (ASes) Announce Prefixes?}
  \label{sec:how-ases}

We first describe how different ASes announce anycast prefixes, quantify prefixes and deployment locations per AS, and distinguish high and low performing prefixes within large anycast providers.

\subsection{Anycast prefixes by Autonomous Systems (ASes)}
  \label{sec:anycast-prefix-by-ases}

\begin{figure}[t]
  \centering

  \begin{subfigure}[t]{\columnwidth}
    \centering
    \includegraphics[width=\columnwidth]{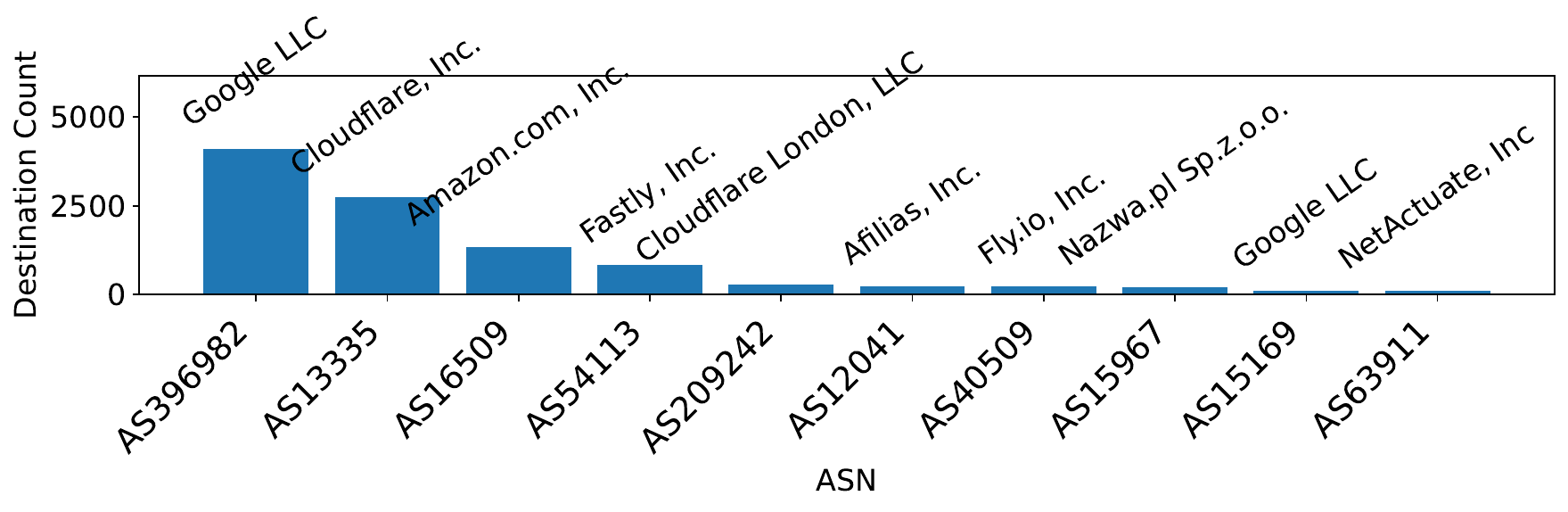}
    \caption{Top 10 ASNs with the number of destination prefixes}
    \label{fig:top-10-asns}
  \end{subfigure}

  \shortpaper{\vspace{0.5em}}

  \begin{subfigure}[t]{\columnwidth}
    \centering
    \includegraphics[width=\columnwidth]{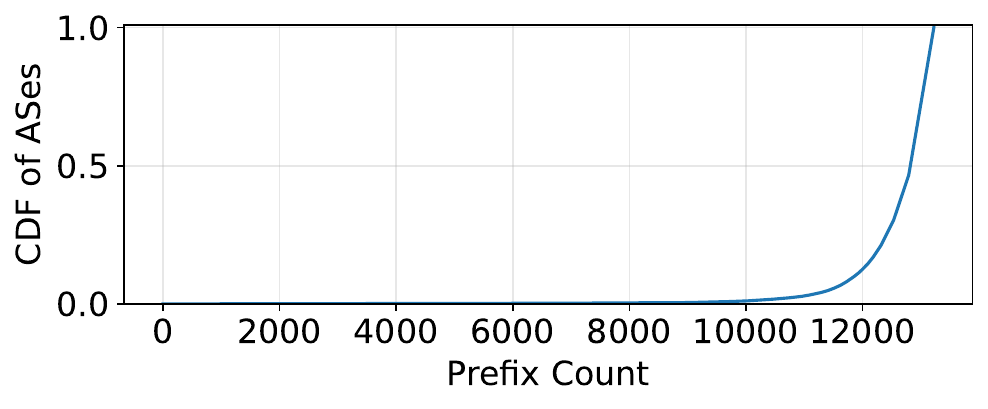}
    \caption{CDF of ASes }
    \label{fig:prefix-per-asn}
  \end{subfigure}

  \caption{Impact of large ASns on number of prefixes}
  \label{fig:asns-prefix}
\end{figure}

Autonomous Systems (ASes) announce prefixes
  for different services like Domain Name System (DNS) 
  or Content Delivery Networks (CDNs). 
Some ASes announce a 
  large number of anycast prefixes. 
The anycast census data indicates that 
  808 ASes announce approximately 13{,}000 anycast prefixes.
The next question is: how many prefixes does each AS announce?

\autoref{fig:asns-prefix} illustrates how large ASes
  influence the distribution of anycast prefixes.
\autoref{fig:top-10-asns} lists the top 10 ASes; 
  the top three (Google, Cloudflare, and Amazon) 
  together account for over 8{,}000 anycast prefixes.
Finally, \autoref{fig:prefix-per-asn} shows that 
  fewer than 5\% of ASes announce more than 10{,}000 anycast prefixes.
After the first 11{,}000 prefixes, 
  the number of ASes rises sharply. 
These additional ASes each announce only a 
  small number of anycast prefixes.

Many prefixes within the same AS are deployed 
  from the same set of anycast sites. 
When these prefixes share the same locations and 
  have similar connectivity to upstream providers and peers, 
  probes are likely to observe 
  similar latencies across those prefixes.
We will see how the performance varies across prefixes within an AS in \autoref{sec:big-ases-latency}.

\subsection{How many sites in different continents?}
  \label{sec:num-sites-across-continents}

\begin{figure}
  \centering
  \includegraphics[width=\columnwidth]{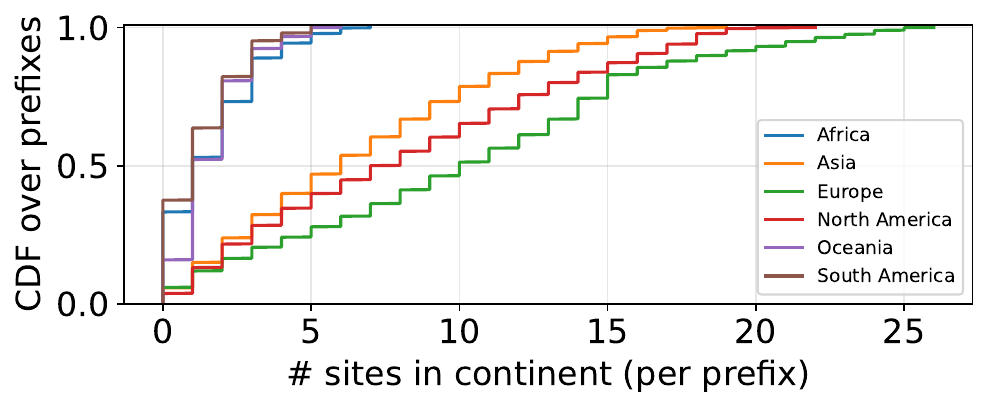}
  \caption{CDF of sites per continent.}
  \label{fig:cdf_continents}
\end{figure}

Anycast providers deploy different numbers of sites 
  across continents depending on their requirements and constraints. 
The number of sites per continent is an important factor 
  when comparing anycast performance across regions.

We examine the number of sites on each continent 
  to understand how anycast deployments vary across regions.
\autoref{fig:cdf_continents} shows the distribution
  of number of sites across continents 
  using the anycast census data~\cite{manycast}.
Oceania, Africa, and South America differ substantially 
  from the other continents in terms of site counts. 
Specifically, 18\% of prefixes have no sites in Oceania, 
  while 25\% and 40\% of prefixes have no sites 
  in Africa and South America, respectively.

\begin{figure}
  \centering
  \includegraphics[width=\columnwidth]{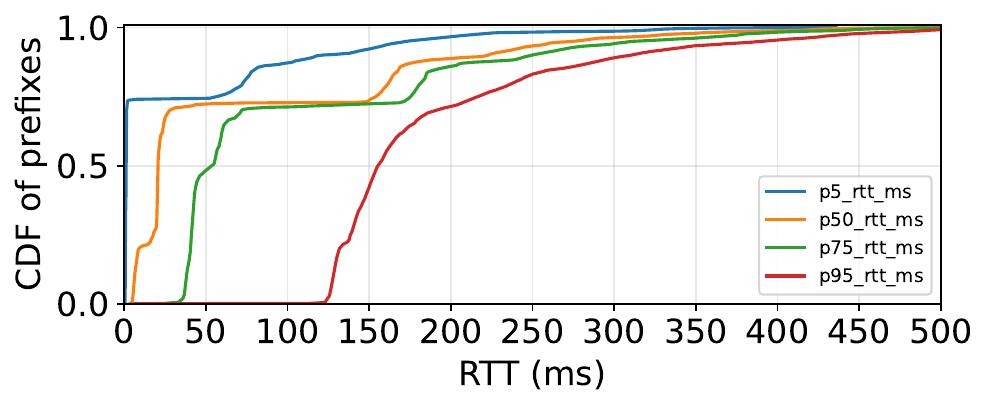}
  \caption{Best case, median, and worst case latency across anycast prefixes (using probes from Africa)}
  \label{fig:cdf_latency}
  \shortpaper{\vspace{-10pt}}
\end{figure}

\subsection{Impacts of number of sites on latency}
  \label{sec:impacts-latency}

We observe that Africa, Oceania, and South America 
  host fewer anycast sites than the other continents. 
Next, we examine how the number of sites affects latency. \autoref{fig:cdf_latency} shows the latency from probes 
  in Africa to all anycast prefixes. 
For each destination address, 
  we compute the best-case ($5^{th}$percentile), median, and worst-case latency ($75^{th}$ and $95^{th}$ percentile). 
For about 75\% of prefixes, at least 5\% African probes observe 
  very low latency because they have a nearby site (blue line in \autoref{fig:cdf_latency}). 
This aligns with our earlier finding 
  that 25\% of anycast prefixes have no sites in Africa (\autoref{sec:num-sites-across-continents}).
For 75\% of prefixes, the median latency is 
  below 50\,ms, indicating that when sites are 
  present in Africa, most probes experience low latency. 
However, the 75th-percentile curve (green line) shows that 
  for more than 50\% of prefixes, 
  at least 25\% of probes observe latencies above 50\,ms.
This result shows that nearby sites matter, but even when a nearby site exists, clients may still experience poor latency.

\subsection{Big ASes with multiple prefixes}
  \label{sec:big-ases-latency}


\begin{figure*}[t]
  \centering

  \begin{subfigure}[t]{\textwidth}
    \centering
    \includegraphics[width=\textwidth]{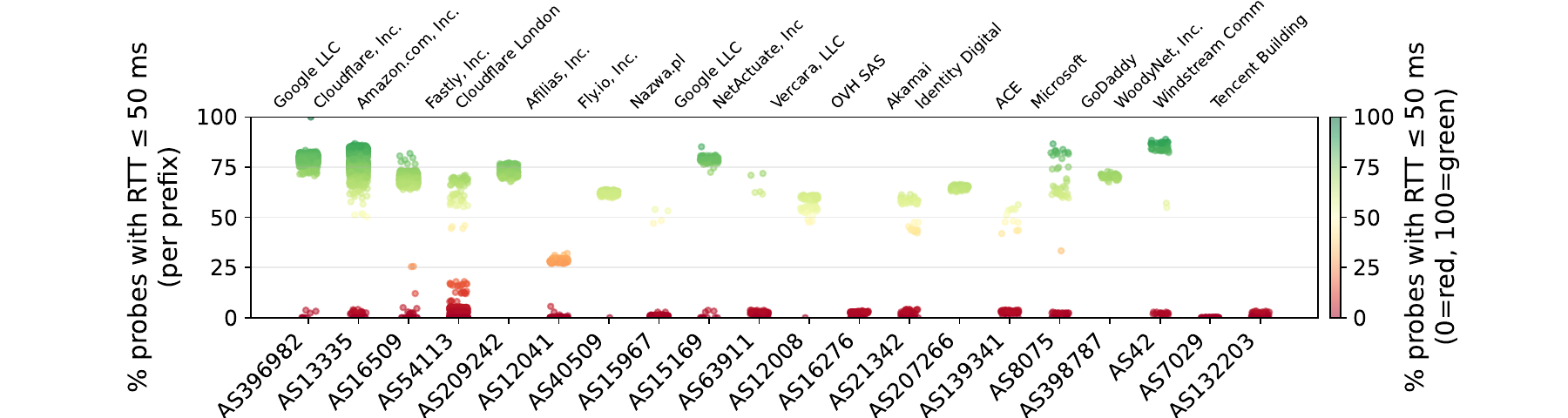}
    \caption{Top 20 ASes and their latency distribution across prefixes}
    \label{fig:top-20-as-latency-distribution}
  \end{subfigure}

  \shortpaper{\vspace{0.5em}}

  \begin{subfigure}[t]{\textwidth}
    \centering
    \includegraphics[width=\textwidth]{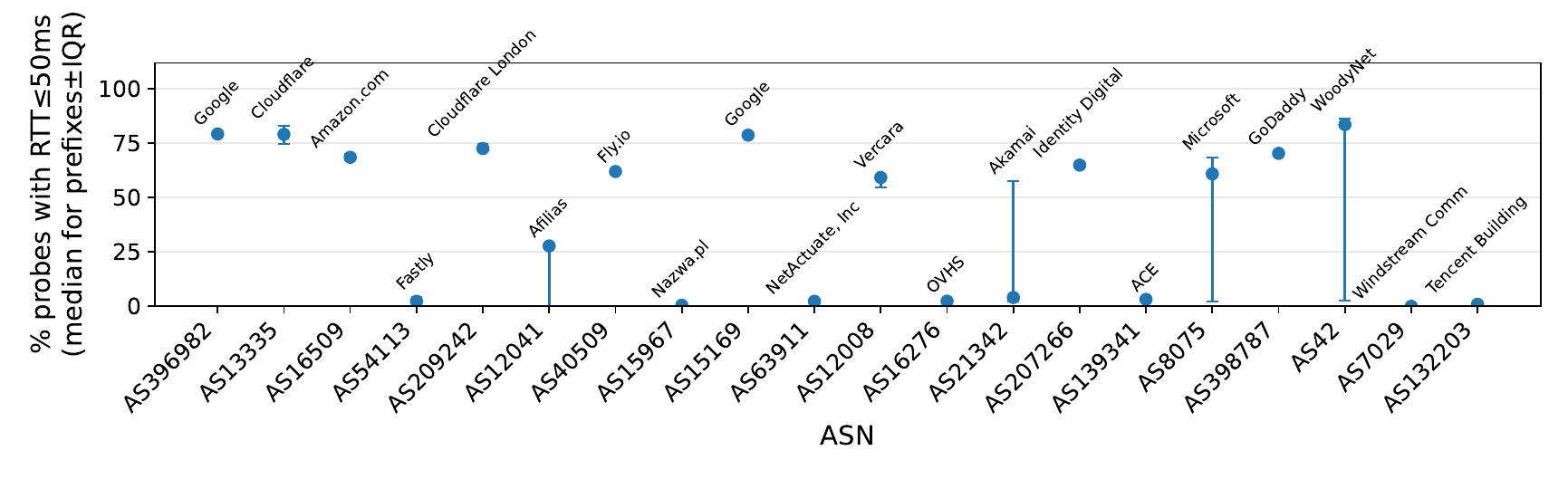}
    \caption{Top 20 ASes and the median percentage of probes with good latency across anycast prefixes}
    \label{fig:top-20-as-median}
  \end{subfigure}

  \caption{Top 20 ASes and their latency distribution based on the vantage points from Africa}
  \label{fig:top-20-asns-latency}
  \shortpaper{\vspace{-4pt}}
\end{figure*}

We investigate how large ASes that announce 
  many anycast prefixes configure their deployments. 
Do they ensure that all prefixes achieve similar latency, 
  or do they configure different prefixes differently? 
This question is important because clients can be served using the prefixes that perform better than others.
If clients or DNS resolvers can choose from multiple prefixes, they can select the best-performing one.
We evaluate this question for the top 20 ASes 
  by number of prefixes.
All these prefixes announce more than 50 prefixes (\autoref{fig:top-10-asns} shows top 10).

\autoref{fig:top-20-asns-latency} shows 
  the latency distribution across prefixes 
  for the top 20 ASes. 
For each prefix, we compute the fraction of probes from Africa (due to space constraints, we present this analysis only for Africa) 
  that observe good latency (RTT $\le$ 50\,ms).
We report this percentage for each anycast prefix announced by that AS (\autoref{fig:top-20-as-latency-distribution}).
We use a red-to-green color scale, 
  where red indicates a lower fraction of probes experiencing 
  good latency and green indicates a higher fraction (\autoref{fig:top-20-as-latency-distribution}).
We also take the median of this fraction 
  across prefixes within each AS (\autoref{fig:top-20-as-median}). 
We report the interquartile range (IQR) 
  across prefixes to capture how this fraction 
  varies across an AS’s prefixes.

Five of the top 20 ASes have no prefix that show good latency for more than 50\% probes (no green data points in \autoref{fig:top-20-as-latency-distribution}). 
Afilias Inc, Nazwa.pl, OVH SAS, Windstream Communications, and 
  Tencent Building are examples 
  of top anycast ASes for which only 
  a small fraction of probes observe good latency.
According to the anycast census data, 
  these five ASes have no anycast sites in Africa, 
  leading to higher overall latency~\cite{manycast}.
Google, Cloudflare, Amazon, Microsoft, and WoodyNet 
  each have multiple prefixes for which 
  roughly 80\% of African probes observe good latency.
According to the anycast census data~\cite{manycast}, 
  Google has at least four anycast sites in Africa, 
  Cloudflare has five, Amazon has three, and WoodyNet has seven, 
  whereas Microsoft has no anycast sites on the Africa continent according to anycast census data. 
Despite this, some Microsoft prefixes still provide 
  good latency to many African probes. 
To investigate, we run traceroutes from a 
  subset of African probes to these prefixes and 
  examine the penultimate hop within Microsoft’s network. 
We observe that a South Africa probe reaches Microsoft 
  in Johannesburg (e.g., ae85-0.ier01.jnb02.ntwk.msn.net), indicating local ingress into Microsoft’s backbone. 
In contrast, a Tunisia probe is routed to 
  Marseille (e.g., ae102-0.icr02.mrs21.ntwk.msn.net), suggesting that 
  some Northern African traffic is served via nearby European site while still achieving low latency due to geographic proximity.

Some ASes configure only a small subset of their prefixes 
  to provide good latency to African probes. 
To quantify this, we compute, for each AS, 
  the median fraction of its anycast prefixes that achieve good latency, along with the IQR across prefixes (\autoref{fig:top-20-as-median}).
Eight of the top 20 ASes exhibit very 
  low median values (points in \autoref{fig:top-20-as-median}), 
  including Fastly, Nazwa.pl, NetActuate, OVH SAS, Akamai, ACE, Windstream Communications, and Tencent Building. 
For these ASes, most announced prefixes deliver 
  poor latency to African probes.

Most ASes configure their anycast prefixes heterogeneously: 
  within the same AS, some prefixes provide good latency 
  while others perform poorly. 
This produces a clear split between green and red points in \autoref{fig:top-20-as-latency-distribution}. 
A smaller set of ASes exhibit more homogeneous configurations, 
  where nearly all prefixes show similar latency behavior. 
For example, Fly.io, Vercara, and Identity Digital display 
  broadly consistent latency distributions across their prefixes.

This performance gap suggests that, even within the same provider, some anycast prefixes outperform others. Therefore, when multiple prefixes offer the same service, selecting the right prefix is important for achieving better performance.

\subsection{Real clients behind prefixes}
  \label{sec:real-clients-measurement}

In \autoref{sec:big-ases-latency}, we showed that 
  large ASes often announce multiple prefixes and 
  that performance can vary substantially across those prefixes.
In particular, some prefixes exhibit poor performance 
  in certain geographic regions. 
This raises a natural follow-up question: 
  \textbf{are clients in those regions actually being served by (or querying for) the poorly performing prefixes?}

To determine whether clients from a given geographic region 
  use these prefixes, we analyze DNS data to verify 
  whether clients in that region issue DNS queries 
  associated with them. 
We first perform reverse DNS lookups to obtain 
  hostnames for the anycast IP addresses. 
We then use B-root DNS data to check whether 
  these hostnames appear in observed DNS queries. 
When a hostname is found, we extract the corresponding 
  source IP addresses from the DNS logs and geolocate them. 
This allows us to identify the geographic regions 
  from which the anycast hostnames were queried, and 
  thus assess whether the relevant prefixes are being 
  used by clients in those regions.

We selected 33 hostnames across six providers and 
  analyzed five hours of DNS data.
We chose hostnames from each AS according to their 
  performance groups: up to three hostnames from the 
  well-performing group and three from the poorly performing group.
These providers were chosen from the top 10 ASes ranked 
  by number of announced prefixes. 
We did not include all top-10 providers because, 
  for some of them, we were unable to obtain hostnames 
  for every anycast prefix. 
We then searched the DNS logs for queries to 
  these hostnames to identify the corresponding 
  source IP addresses and infer their geographic locations.

We used hostnames from six ASes—WoodyNet, Afilias, 
  NetActuate, Amazon, Google, and Akamai—and selected 
  multiple prefixes from each AS representing both best-case 
  and worst-case latencies. 
In the B-root DNS logs, we were able to identify 
  hostnames from both best and worst performing groups associated with WoodyNet, Afilias, NetActuate, 
  and Amazon. 
We did not observe hostnames for Google or Akamai because we analyzed only five hours of data; additionally, client traffic may have been directed to other root servers, and some clients may have been served from cache.
This result shows that African clients query prefixes from 
  both the best-performing and worst-performing latency groups.

\section{How is the polarization?}
  \label{sec:polarization}

Next, we present a comparative analysis of polarization and performance of the anycast networks.

  
  


\iftoggle{asia}
{

\begin{table*}[t]
\centering
\small
\setlength{\tabcolsep}{1.6pt}
\renewcommand{\arraystretch}{1.05}

\newcolumntype{P}[1]{>{\raggedright\arraybackslash}p{#1}}

\resizebox{\textwidth}{!}{%
\begin{tabular}{|c| l P{2.0cm} >{\columncolor{green!12}}r |
                c| l P{2.0cm} >{\columncolor{yellow!18}}r |
                c| l P{2.0cm} >{\columncolor{red!12}}r |
                c| l P{2.0cm} >{\columncolor{red!12}}r|}
\hline
\multirow{11}{*}{\centering NA}
& ASN & AS name & Min pol (\%)
& \multirow{11}{*}{\centering SA}
& ASN & AS name & Min pol (\%)
& \multirow{11}{*}{\centering AS}
& ASN & AS name & Min pol (\%)
& \multirow{11}{*}{\centering AF}
& ASN & AS name & Min pol (\%) \\
\hline
& AS396982 & Google      & 1.1 & & AS54113  & Fastly            & 4.4 & &  AS13335    & Cloudflare   & 11.9   & & AS21556 & NASA          & 12.3 \\
& AS15169  & Google      & 1.2 & & AS13335  & Cloudflare        & 4.9 & & AS209242     & Cloudflare  & 12.6    & & AS1335  & Cloudflare    & 13.6 \\
& AS16509  & Amazon      & 1.2 & & AS15169  & Google            & 6.5 & & AS398787     & GoDaddy  &  12.7   & & AS42    & WoodyNet      & 13.7 \\
& AS13335  & Cloudflare  & 1.2 & & AS396982 & Google            & 6.9 & &        &                   &     & & AS15169 & Google        & 15.2 \\
& AS54113         & Fastly            & 1.8    & & AS209242 & Cloudflare & 7.1 & & AS23467   &  New Relic                 &  13.3   & & AS10886 & U-of-Maryland & 15.5 \\
&  AS398787	 &  GoDaddy  & 1.8    & & AS398787 & GoDaddy           & 7.5 & & AS19254 &  Koch      &     13.4          & & AS187   & DNScast       & 15.9 \\
&  AS2635	 & Automattic     & 2.3     & & AS55532  & Squiz             & 7.6 & & AS55532   & Squiz Pty   & 13.5    & & AS8075  & Microsoft     & 17.3 \\
&  AS55002	& F5 &  2.3   & & AS62659  & Q2 Soft           & 7.9 & & AS4749 &  WiseTech      &  13.6        & & AS19281 & Quad9         & 17.6 \\
&  AS19905	   &  Vercara  & 2.3    & & AS53332  & C2Hosting         & 8.2 & & AS50756       & Mikrofinanso  & 13.9     & & AS396982& Google        & 17.7 \\
&  AS29085	  & Porsche  & 2.4    & & AS2559   & Visa              & 8.3 & & AS399566  & Bigcommerce   & 14.0    & & AS8068  & Microsoft     & 17.9 \\
\hline
\end{tabular}%
}

\caption{Top ASes with lowest polarization in North America (NA), South America (SA), Asia (AS), and Africa (AF).}
\label{tab:as-min-polarization}
\shortpaper{\vspace{-10pt}}
\end{table*}

}
{
\begin{table}[t]
\centering
\small
\setlength{\tabcolsep}{1.8pt}      
\renewcommand{\arraystretch}{1.0}

\resizebox{\columnwidth}{!}{%
\begin{tabular}{|c| l  p{1.0cm}  r | c | l  p{1.0cm}  r|}
\hline
\multirow{11}{*}{\centering AF}
& ASN & AS name & Min pol (\%)
& \multirow{11}{*}{\centering SA}
& ASN & AS name & Min pol (\%) \\
\hline
& AS21556 & NASA               & 12.3 & & AS54113  & Fastly             & 4.4 \\
& AS1335  & Cloudflare         & 13.6 & & AS13335  & Cloudflare         & 4.9 \\
& AS42    & WoodyNet           & 13.7 & & AS15169  & Google             & 6.5 \\
& AS15169 & Google             & 15.2 & & AS396982 & Google             & 6.9 \\
& AS10886 & U-of-Maryland      & 15.5 & & AS209242 & Cloudflare-London  & 7.1 \\
& AS187   & DNScast            & 15.9 & & AS398787 & GoDaddy            & 7.5 \\
& AS8075  & Microsoft          & 17.3 & & AS55532  & Squiz              & 7.6 \\
& AS19281 & Quad9              & 17.6 & & AS62659  & Q2 Soft            & 7.9 \\
& AS396982& Google             & 17.7 & & AS53332  & C2Hosting          & 8.2 \\
& AS8068  & Microsoft          & 17.9 & & AS2559   & Visa               & 8.3 \\
\hline
\end{tabular}%
}

\caption{Top ASes with lowest polarization (best case) in AF and SA.}
\label{tab:as-min-polarization}
\shortpaper{\vspace{-10pt}}
\end{table}
}

\full{
\begin{figure*}[t]
  \centering
  \shortpaper{\vspace{-4pt}}

  \newcommand{\mapheight}{5.0cm}

  \begin{subfigure}[t]{0.33\textwidth}
    \centering
    \includegraphics[height=\mapheight,keepaspectratio]{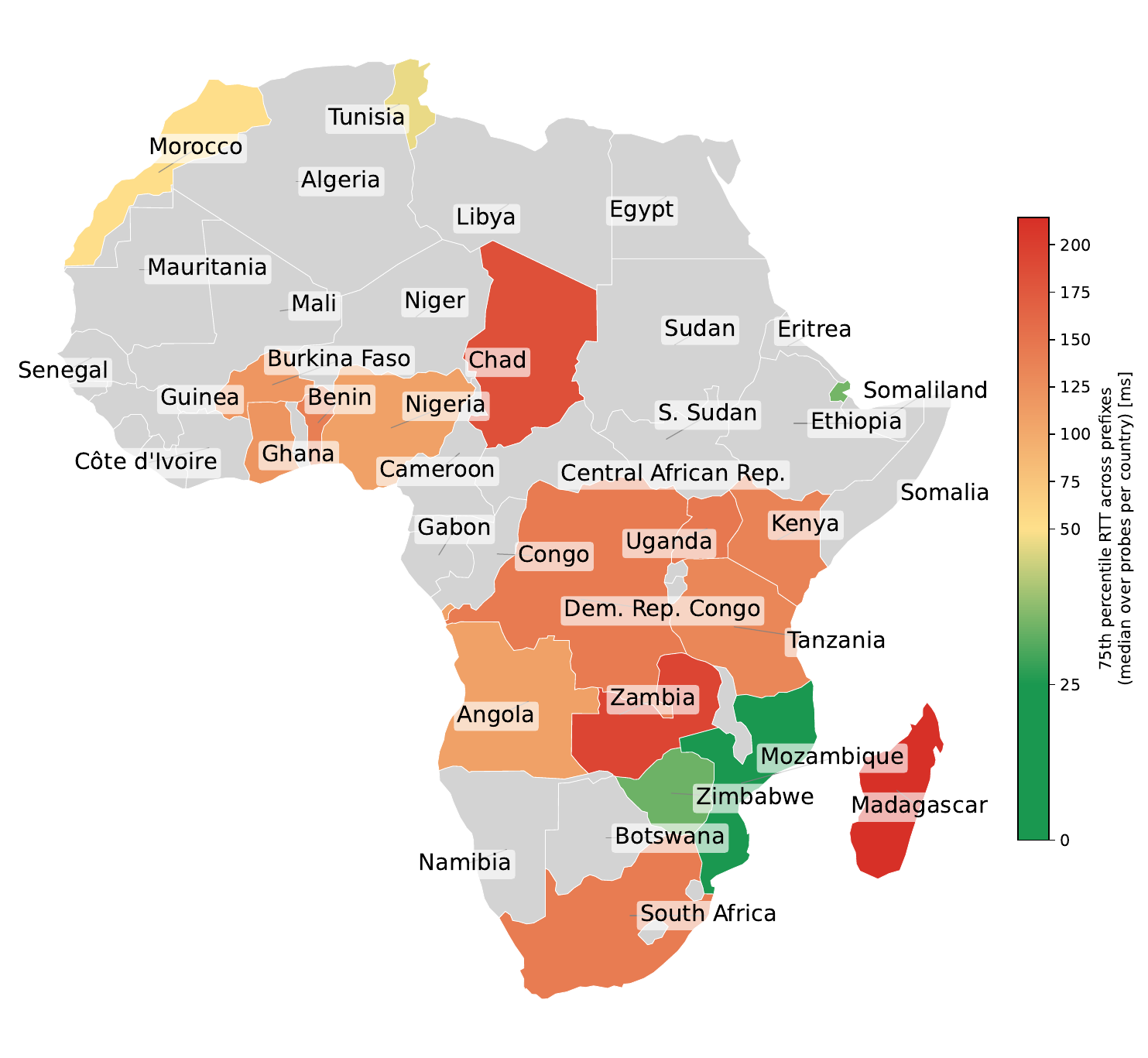}
    \caption{Africa.}
    \label{fig:africa-75}
  \end{subfigure}\hfill
  \begin{subfigure}[t]{0.33\textwidth}
    \centering
    \includegraphics[height=\mapheight,keepaspectratio]{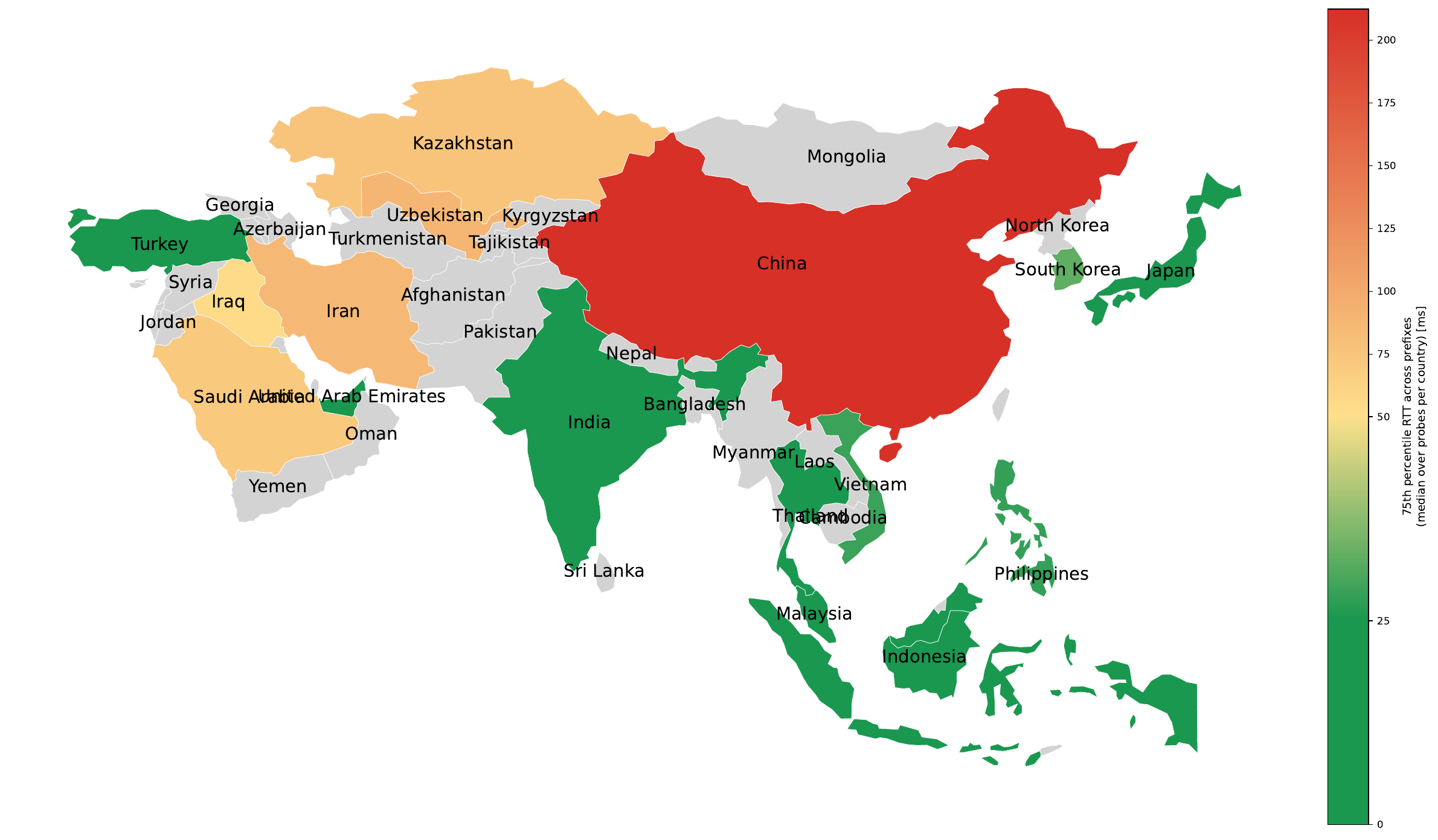}
    \caption{Asia.}
    \label{fig:asia-75}
  \end{subfigure}\hfill
  \begin{subfigure}[t]{0.33\textwidth}
    \centering
    \includegraphics[height=\mapheight,keepaspectratio]{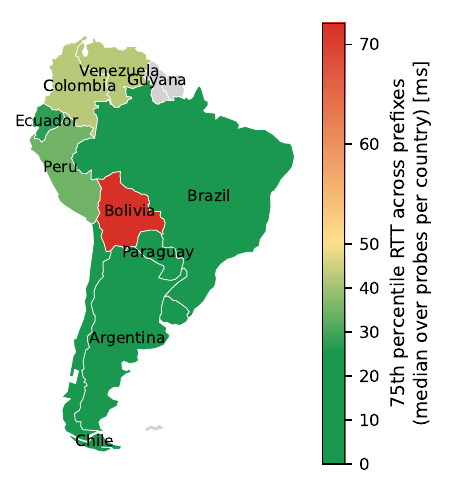}
    \caption{South America.}
    \label{fig:sa-75}
  \end{subfigure}

  \shortpaper{\vspace{-4pt}}
  \caption{Country-level median latency variation in Africa, Asia, and South America ($75^{th}$ percentile latency across prefixes).}
  \label{fig:africa_sa_asia_latency_maps}
  \shortpaper{\vspace{-6pt}}
\end{figure*}

}

\begin{figure}
  \centering
  \includegraphics[width=\columnwidth]{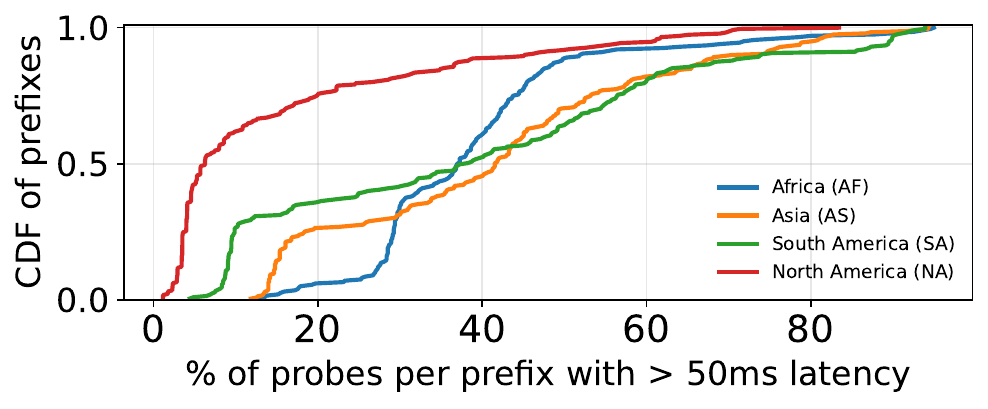}
  \caption{CDF of polarization level across anycast prefixes in Africa, \asia{Asia,} North America, and South America}
  \label{fig:cdf-polarization-africa}
\end{figure}

Polarization occurs when probes from the same geographic region observe markedly 
  different performance---some receive good latency while others experience poor latency. 
For a given continent, 
  we define a prefix as polarized if more than 5\% of probes 
  observe good latency (RTT $\le$ 50\,ms) while the remaining probes observe poor latency. 
In Africa, 75.7\% of prefixes (9,831 out of 12,993) have 
  at least 5\% of probes with RTT $\le$ 50\,ms. 
For the remaining 24.3\% of prefixes, most probes experience 
  poor latency, which we suspect is due to the 
  absence of anycast sites in Africa or they are local anycast prefixes for another continent. 
Moreover, among the prefixes where at least 5\% of probes 
  observe good latency, we find that all of them 
  exhibit polarization.

Probes from different continents exhibit 
  distinct latency distributions. 
\autoref{fig:cdf-polarization-africa} shows the 
  latency distribution for probes from four continents. 
As expected, North America has substantially 
  lower polarization than Africa\asia{, Asia,} and South America. 
For probes in Africa, for 20\% of anycast prefixes, 
  at least 30\% of African probes observe RTTs above 50\,ms.
For 75\% of prefixes, at least 45\% of African probes 
  observe RTTs above 50\,ms. 
\asia{Asia exhibits a better best-case scenario than Africa; however, for over 50\% of prefixes, more than 40\% of probes experience latencies exceeding 50\,ms.}
South American probes show lower polarization 
  for roughly 45\% of prefixes, 
  while for the remaining prefixes Africa performs better.
North American probes perform markedly better than 
  both Africa, Asia and South America: 60\% of prefixes exhibit 
  less than 10\% polarization, which is even better 
  than the best-case scenario (12.3\% minimum polarization) observed in Africa.
\autoref{fig:cdf-polarization-africa} shows how polarization differs across prefixes announced by different ASes. 
Across all continents, we observe a wide range of polarization levels rather than a single consistent pattern.

\textbf{Best performing anycasts:}
\autoref{tab:as-min-polarization} lists the ASes with the lowest polarization observed for African and South American probes. The smallest minimum polarization we find is 12.3\% for AS21556 (NASA) in Africa, indicating that even for its best-performing prefix, 12.3\% of African probes still see RTTs above 50\,ms despite the presence of nearby anycast sites. Other ASes with similarly low minimum polarization include WoodyNet, the University of Maryland, Cloudflare, DNScast, Microsoft, Google, Quad9, and Amazon for African probes. For these ASes, the prefixes that perform the best exhibit minimum polarization values between 12\% and 18\%. 
\asia{The best-case scenario in Asia is slightly better than Africa---still showing 11.9\% to 14.0\% polarization.}
South American probes show a better best-case outcome, with minimum polarization as low as 4.4\%, achieved by Fastly.
Many South American probes are routed to anycast sites in North America that are geographically relatively close, leading to lower latency.

\full
{
\textbf{Latency variation across countries:}
Countries in Africa and South America experience different 
  latency levels to anycast prefixes. 
For each country, we first compute 
  the median RTT across all probes in that country 
  for each prefix; we then report the 75th-percentile RTT across prefixes. 
We cannot collect measurements for countries without RIPE Atlas probes, and we include only countries with at least three probes.

Many African countries exhibit median latencies around 100\,ms from their probes to at least 25\% of the prefixes (e.g., Egypt, South Sudan, Uganda, and Kenya). In more extreme cases, some countries experience median latencies above 150\,ms to 25\% of the anycast prefixes (e.g., Chad, Cameroon, and Congo). Overall, southern Africa (e.g., South Africa and Zimbabwe) tends to perform better than central regions of the continent.
\full{\autoref{fig:africa-75} shows the latency distribution across countries in Africa.}

In Asia, multiple countries—including India, South Korea, Japan, and parts of Southeast Asia—exhibit good median latency ($\sim$25\,ms) to at least 75\% of the prefixes \full{(\autoref{fig:asia-75})}. In contrast, Middle Eastern countries such as Iran and Saudi Arabia show median latencies of 50–75\,ms, while China exceeds 150\,ms.

In South America, latency distributions are generally better than in Africa. Even so, probes in Brazil, Paraguay, Argentina, and Chile typically achieve lower latencies than those in Bolivia, Peru, Colombia, and Venezuela. 
\full{We can see country-wise latency map in \autoref{fig:sa-75}.} 
}
 

\section{Reasons behind polarization}
  \label{sec:reasons}

Next, we make a root-cause investigation of the factors driving the observed performance issues and polarization.

\subsection{Why Do Even the Best-Performing Anycast Prefixes Exhibit Polarization?}
  \label{sec:conf-low-polarization}

\begin{figure*}[t]
  \centering

  \begin{subfigure}[t]{\textwidth}
    \centering
    \includegraphics[width=\textwidth]{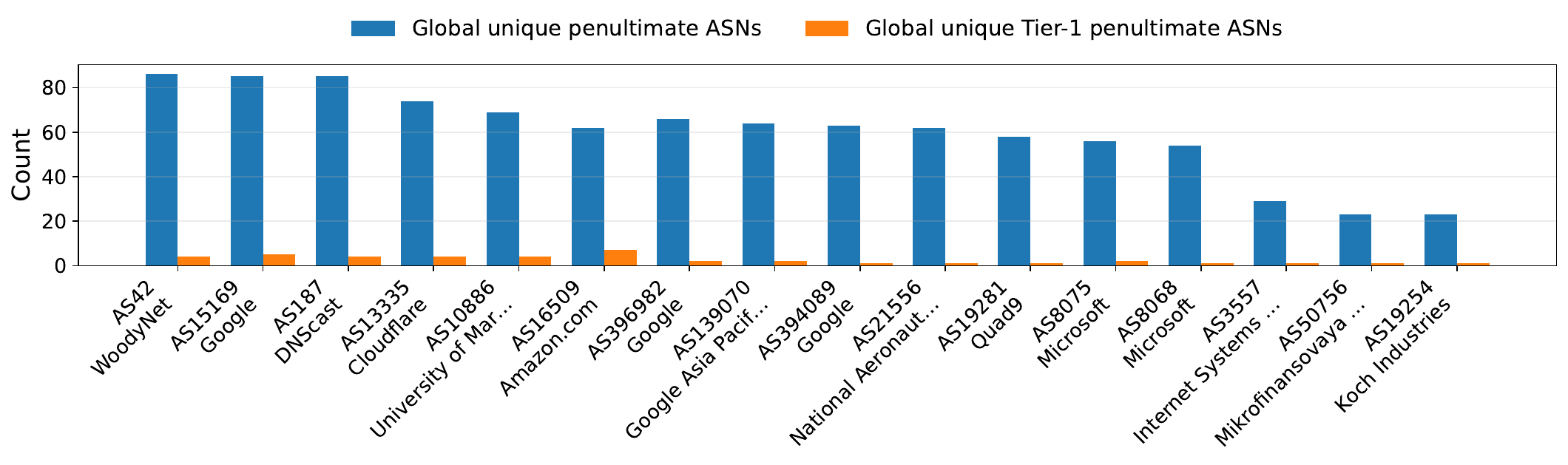}
    \caption{Number of unique penultimate ASNs with Tier-1 count}
    \label{fig:unique-asns}
  \end{subfigure}

  \shortpaper{\vspace{0.5em}}

  \begin{subfigure}[t]{\textwidth}
    \centering
    \includegraphics[width=\textwidth]{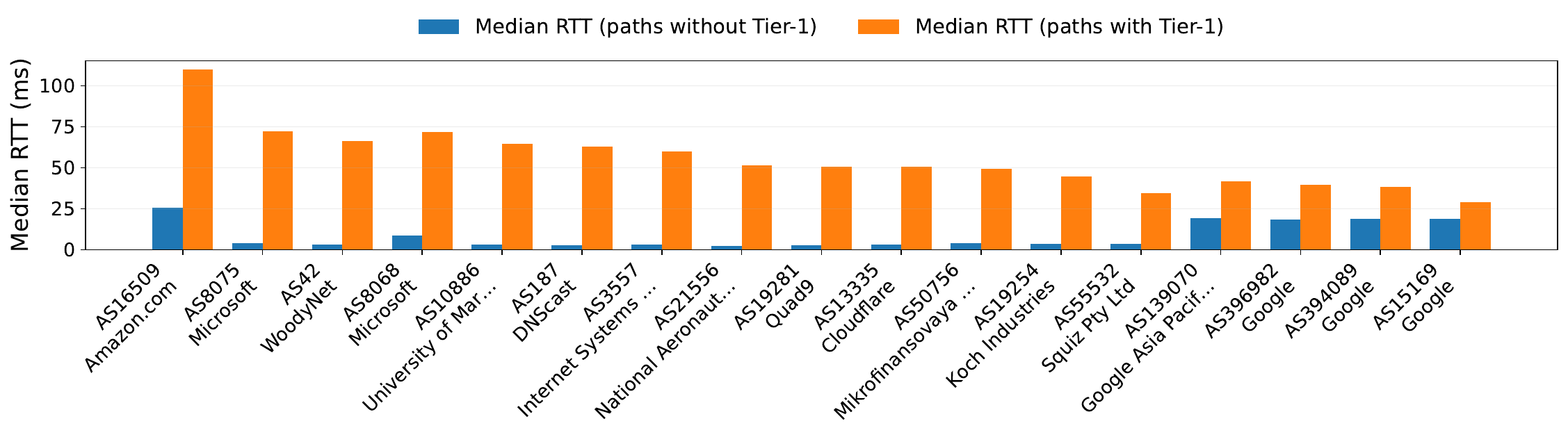}
    \caption{Median RTT comparison when AS paths include Tier-1 and non Tier-1 ASes}
    \label{fig:rtt-w-tier-1}
  \end{subfigure}

  \caption{Number of unique ASes in the routing paths}
  \label{fig:top-20-asns-paths}
\end{figure*}

We examine the \textbf{penultimate AS hops} on paths 
  to the anycast prefixes with the lowest polarization 
  to better understand how these networks are connected. 
We expect to observe different local and regional ASes 
  on the paths that carry traffic from sources to 
  nearby anycast sites. 
To test this, we analyze traceroutes from  
  vantage points of Africa, Asia, and South America 
  to the low-polarization anycast prefixes 
  and extract the corresponding penultimate AS hops. 
We then identify which of these penultimate hops belong to 
  Tier-1 providers to assess 
  (i) how dependent these paths are on Tier-1 transit and 
  (ii) how Tier-1 involvement affects latency.
We define Tier-1 providers as the top 15 ASes 
  in the CAIDA AS rank list (\url{https://asrank.caida.org})
  as these ASes have the largest customer cones.

\autoref{fig:top-20-asns-paths} summarizes the 
  diversity of penultimate AS hops and the 
  latency differences between Tier-1 and non–Tier-1 paths (we only show the results for Africa due to space constraints). 
We first count the number of distinct penultimate ASes 
  observed for each destination and 
  then determine how many of them are Tier-1 ASes. 
\autoref{fig:unique-asns} shows the number of 
  distinct ASes appearing as penultimate hops: 
  some destinations are reached through more than 
  80 distinct penultimate ASes, while the least diverse case 
  has about 20. 
Most destinations fall around 60 distinct 
  penultimate ASes, indicating substantial heterogeneity in 
  how African networks reach these anycast sites.
Although the paths include relatively few distinct Tier-1 ASes, these ASes have a substantial impact on latency.

Latency also differs markedly depending on whether 
  paths traverse Tier-1 providers. 
When traffic is routed through Tier-1 ASes, 
  median latency is consistently higher, 
  reaching over 100\,ms in the worst case (orange bars in \autoref{fig:rtt-w-tier-1}). 
In contrast, paths that avoid Tier-1 provider typically 
  exhibit much lower median latency—often below 
  5\,ms---suggesting delivery to a nearby (local) anycast site (blue bars in \autoref{fig:rtt-w-tier-1}).

\begin{figure}
  \centering
  \includegraphics[width=\columnwidth]{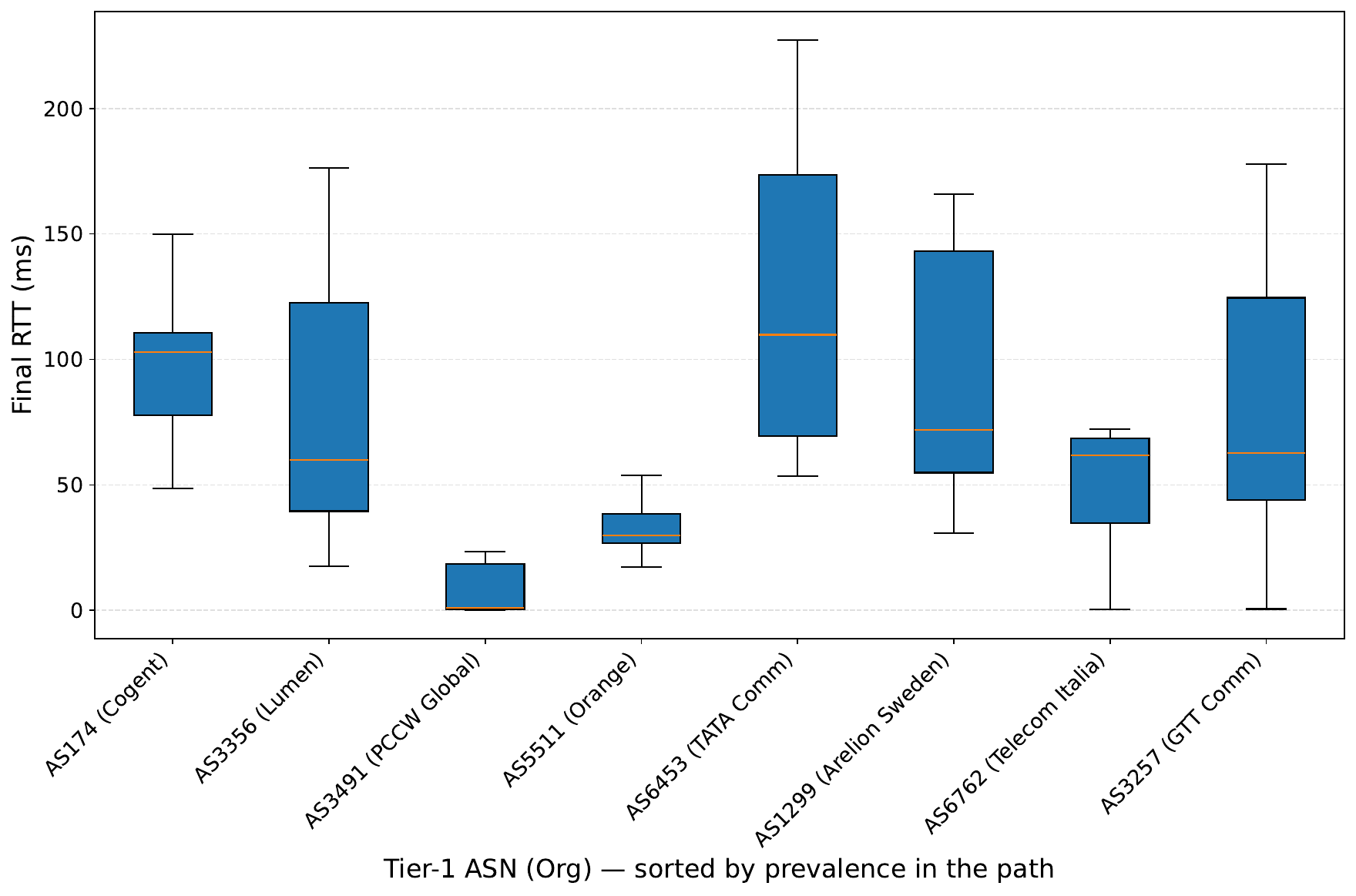}
  \caption{Latency distribution for Tier-1 ASes when the Tier-1 ASes are in the routing path}
  \label{fig:tier-1-latency-box}
  \shortpaper{\vspace{-10pt}}
\end{figure}

\begin{figure*}[t]
  \centering

  \begin{subfigure}[t]{\textwidth}
    \centering
    \includegraphics[width=\textwidth]{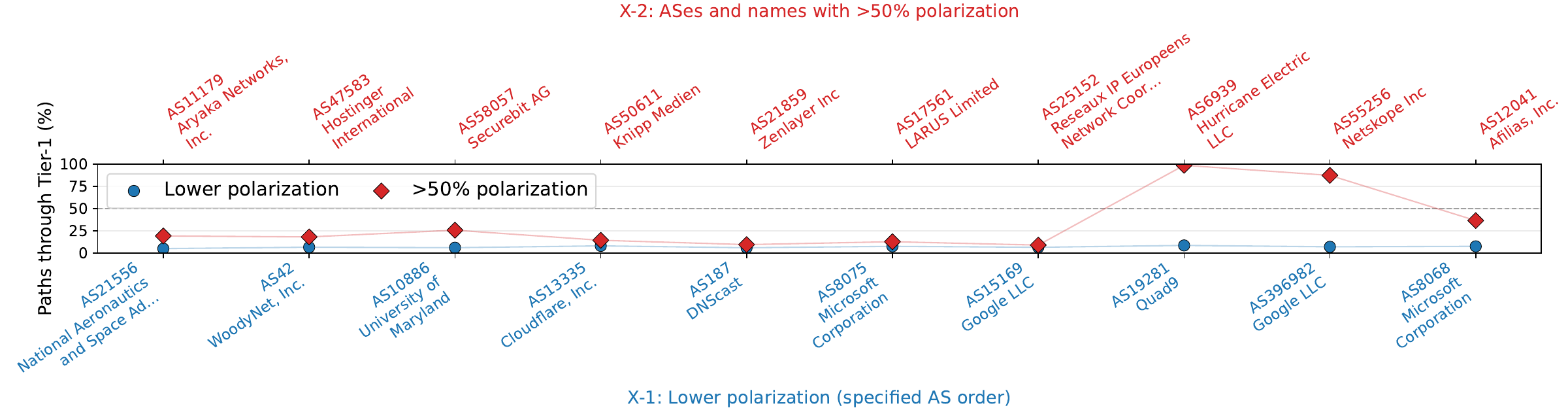}
    \caption{Percentage of Tier-1 AS in the paths}
    \label{fig:pct-tier-1-path}
  \end{subfigure}

  \shortpaper{\vspace{0.5em}}

  \begin{subfigure}[t]{\textwidth}
    \centering
    \includegraphics[width=\textwidth]{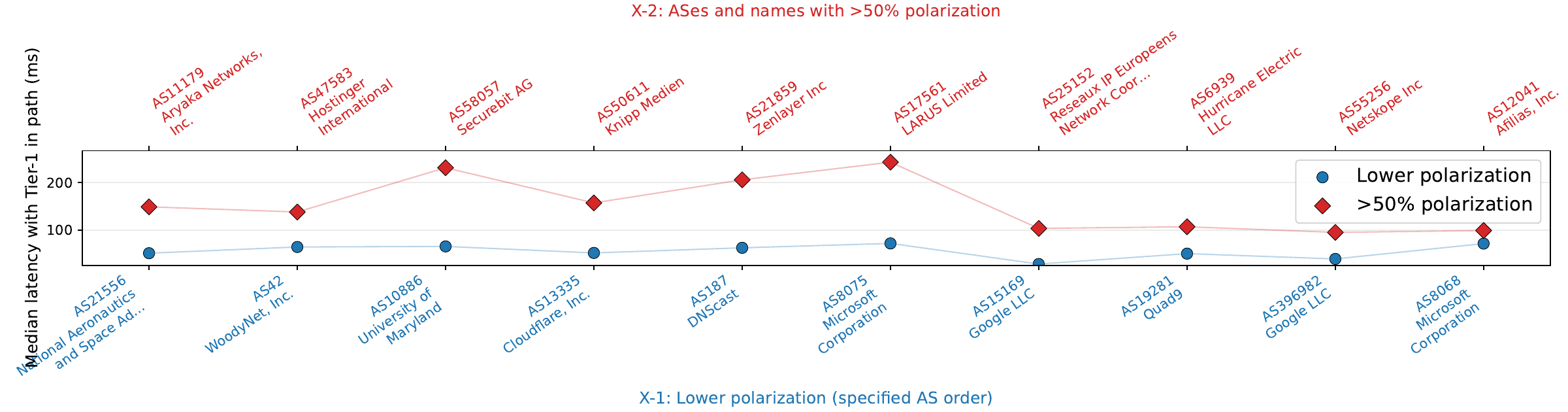}
    \caption{Median latency variation}
    \label{fig:lat-tier-1-comparison}
  \end{subfigure}

  \caption{Comparison of Tier-1 existence and latency variation between lower and $>50\%$ polarization}
  \label{fig:polarization-comparison}
  \shortpaper{\vspace{-10pt}}
\end{figure*}

\textbf{Which Tier-1 ASes are more active?:}
We examined the most active Tier-1 ASes observed in 
  African routes, since their presence in end-to-end paths 
  is often associated with increased latency. 
Our goal is to characterize the routing paths 
  these ASes take and investigate 
  why they exhibit poor latency performance.

AS174 (Cogent Communications) appears most frequently 
  on paths to our destinations. 
Other prominent Tier-1 networks include 
  AS3491 (PCCW Global), AS3356 (Level 3/Lumen), and 
  AS5511 (Orange). 

Latency varies substantially depending on 
  which Tier-1 AS is present along the path.
\autoref{fig:tier-1-latency-box} summarizes the 
  latency distributions, with the x-axis ordered by 
  how prevalent each Tier-1 AS is in the observed paths. 
The most prevalent network, AS174 (Cogent), 
  exhibits a median latency above 100\,ms when 
  it appears on the path. 
AS3356 (Lumen) also shows a broad latency range where
  there is significant gap between the median and $75^{th}$
  percentile latency. 
In contrast, AS3491 (PCCW Global) and AS5511 (Orange) 
  have comparatively stable latency distributions: 
  the worst-case latency remains within 25\,ms for AS3491 
  and within 50\,ms for AS5511.

These large ASes sometimes steer traffic to 
  a distant anycast site because they lack 
  connectivity to the nearest anycast location. 
For instance, when traffic traverses AS174 (Cogent), 
  it may be routed from Africa to a 
  European anycast site, since Cogent is connected 
  to the European site and not to a local African site.
This is one of the reasons behind polarization 
  in anycast networks.

We find anycast networks sometimes use multiple 
  big ASes to have complete coverage and redundancy 
  as shown in \autoref{fig:unique-asns}. 
However, having multiple ASes in the paths 
  may have adverse impacts on latency if 
  all the big transits are not connected 
  locally to an anycast site.
While it is hard to predict how many big ASes
  should be there as the transit providers, 
  it is essential to have complete connectivity 
  with the transit providers.
If the transit is only connected to a few anycast locations,
  then there is a high chance of having polarization.

\textbf{Characteristics in Asia and South America:}
We also assess latency variation in Asia and South America by comparing paths that include a Tier-1 AS with those that do not. We observe similar trends: when Tier-1 ASes appear in the path, packets often take a longer route. In the best-case scenario, paths traversing a Tier-1 AS mostly show a median latency of over 50\,ms for Asia and 30\,ms for South America, whereas paths without a Tier-1 AS have a median latency of under 10\,ms in both Asia and South America. 

AS4637 (Telstra Global), AS9002 (RETN Limited), AS3491 (PCCW Global), AS3356 (Lumen), and AS1299 (Arelion) are the most common Tier-1 ASes in Asia. Among these, AS4637 (Telstra Global) shows a wide range of latency ranging from 5\,ms to 170\,ms. We suspect this AS is locally connected to some anycast prefixes but reaches other prefixes via more distant locations. We recommend anycast providers should establish local connectivity with this provider, as many Asian clients rely on this AS.

We also find that AS3356 (Lumen), AS174 (Cogent), AS1299 (Arelion), and AS6762 (Telecom Italia) are the most frequently observed ASes in paths from South American probes.

\textbf{Other routing inefficiencies:}
To examine how routing inefficiencies can cause poor latency from local ASes, we analyze traceroute data for the best-performing prefixes.
We observe that certain local ASes incur substantial delays before reaching the destination. For example, when AS2609 (Tunisia Backbone) appears as the penultimate AS hop, the median latency is 175\,ms. To understand this behavior, we examined the routing paths traversing AS2609 and found routes to \textit{e.root-servers.net} (AS21556 – NASA). Although this root DNS server has instances in Africa and the corresponding prefix is among the best-performing in the region, paths that traverse AS2609 exhibit significantly higher latency. In these traces, we see multiple routing hops within Tunisia, and the geo-location data suggests an inflated and inefficient routing path.
We also found similar behavior for AS16637 (MTN South Africa), when they carry Cloudflare traffic. 

We also observe cases where local ASes carry traffic to distant regions and effectively behave like transit providers. For instance, AS12956 (Telxius Cable) routes DNSCast traffic from Chad to Europe, resulting in latencies exceeding 100\,ms. Network operators should investigate such routing inefficiencies to identify the cause of elevated delays. In many cases, providers can mitigate the problem by discontinuing prefix announcements through upstream networks that consistently route traffic inefficiently. In Asia, we observed inefficient paths when traffic traverses AS4134 (Chinanet Backbone) to reach a Cloudflare prefix.
When analyzing the best-performing prefixes from South America, we observe similar behavior in a few local ASes.

\subsection{How Do Poorly Performing Anycast Prefixes Differ from the Best-Performing Prefixes?}
  \label{sec:config-variation}

Next, we examine configuration differences under high polarization. Our goal is to understand why certain prefixes exhibit higher polarization and why elevated latency persists across a larger fraction of probes.

We select 10 prefixes with polarization greater than 50\%---that is, more than half of their probes from Africa observe latency above 50\,ms. We then run traceroutes from the same set of probes to these anycast prefixes. For each traceroute, we determine whether the path to the destination includes a Tier-1 AS and identify the penultimate AS hop connected to the anycast site. Finally, we compare the latency distributions of paths that traverse Tier-1 ASes with those that use only non–Tier-1 ASes.

We compare these results with the best-case scenario, which has the lowest polarization (\autoref{fig:polarization-comparison}). Our objective is to identify differences in anycast configuration that contribute to higher polarization. \autoref{fig:pct-tier-1-path} shows the percentage of paths that traverse a Tier-1 AS. In most instances where polarization exceeds 50\%, we observe a higher fraction of Tier-1–transiting paths (the red line lies above the blue line). In some cases, the majority of paths use a Tier-1 AS when polarization is above 50\%. One example is when the anycast deployment belongs to a Tier-1 network itself (AS6939). In other cases, such as AS55256 and AS12041, the anycast deployment appears to rely heavily on Tier-1 ASes. These anycast prefixes depend more on Tier-1 ASes instead of maintaining direct peering relationships with local ASes.

\autoref{fig:lat-tier-1-comparison} compares latency in the lowest-polarization case versus the $>50\%$ polarization case for paths that traverse Tier-1 ASes. We observe higher polarization when performance over Tier-1 paths is significantly worse (the red line sits well above the blue line). In some cases, the median latency exceeds 200\,ms when Tier-1 ASes appear on the path. This suggests these Tier-1 ASes are not well connected to local or nearby anycast sites, forcing traffic onto longer routes to reach the destination and leading to higher latency. 

We also compare the median latency in the lowest-polarization case and the high-polarization case for paths that do not traverse a Tier-1 AS. Even without Tier-1 ASes, the higher-polarization cases still exhibit higher median latency than the best-case scenario. However, when Tier-1 ASes are absent, median latency remains much lower (typically $<50$\,ms) than in the Tier-1–transiting case. Overall, polarization appears to be driven primarily by a greater reliance on Tier-1 transit and by Tier-1 connectivity that steers traffic to distant anycast sites.

When comparing to the best case scenario, we observe many of the same Tier-1 ASes on paths when polarization exceeds 50\%, but their performance is markedly worse in these cases. For instance, AS3491 (PCCW) exhibits a median latency of about 100\,ms, whereas in the best-case scenario it was below 10\,ms. This highlights the importance of strong local connectivity with large transit providers. In some poorly performing anycast prefixes, different large ASes exist in the paths; however, it is generally preferable to have broad connectivity to local anycast sites rather than relying on several Tier-1 ASes that are connected to only a limited subset of sites.

\section{Conclusion}
  \label{sec:conclusion}
In this paper, we evaluate anycast networks to identify polarization in regions such as Africa, Asia, or South America, where Internet connectivity and interconnection ecosystems are comparatively less mature. We analyze more than 10,000 anycast prefixes by probing them from RIPE Atlas vantage points located in Africa, Asia, and South America. Using these measurements, we quantify the extent of polarization in each region. We illustrate how large providers deploy their networks and prefixes, highlighting cases with both low and high latency. We find that polarization is substantially more severe in Africa, Asia, and South America than in North America. To understand the underlying causes, we complement latency measurements with traceroutes to infer anycast configurations and routing behavior, and we show that large ASes frequently steer traffic to distant anycast sites, contributing directly to polarization. We also highlight routing inefficiencies where traffic from a local AS takes a longer path to the destination. Overall, our results suggest that improving the connectivity between large providers and local anycast sites is the key to mitigating polarization.




\bibliographystyle{IEEEtran}  
\bibliography{sample}

\end{document}